\documentclass[epj,nopacs]{svjour}
\usepackage{graphicx}
\graphicspath{{./eps/}}
\newcommand{\NOTE}[1]{}
\renewcommand{\NOTE}[1]{\marginpar{\raggedright \scriptsize \sloppy #1}}
\begin{document}
\hugehead
\title{Scaling of Particle and Transverse Energy Production
       in \boldmath $^{208}$Pb\/+\/$^{208}$Pb collisions at 158$\cdot A$~GeV
       \unboldmath}
\subtitle{WA98 Collaboration}
\author{
     M.M.~Aggarwal\inst{1}
\and A.~Agnihotri\inst{2}
\and Z.~Ahammed\inst{3}
\and A.L.S.~Angelis\inst{4} 
\and V.~Antonenko\inst{5}
\and V.~Arefiev\inst{6}
\and V.~Astakhov\inst{6}
\and V.~Avdeitchikov\inst{6}
\and T.C.~Awes\inst{7}
\and P.V.K.S.~Baba\inst{8}
% ...
}
\institute{
     University of Panjab, Chandigarh 160014, India
\and University of Rajasthan, Jaipur 302004, Rajasthan, India
\and Variable Energy Cyclotron Centre,  Calcutta 700 064, India
\and University of Geneva, CH-1211 Geneva 4,Switzerland
\and RRC ``Kurchatov Institute'', RU-123182 Moscow, Russia
\and Joint Institute for Nuclear Research, RU-141980 Dubna, Russia
\and Oak Ridge National Laboratory, Oak Ridge, Tennessee 37831-6372, USA
\and University of Jammu, Jammu 180001, India
% ...
}
%\authorrunning{Aggarwal et al.}
%
\date{Received: date / Revised version: date}
\date{Draft 1.6, \today}
\date{\today}

\abstract{ Transverse energy, charged particle pseudorapidity
distributions and photon transverse momentum spectra have been
studied as a function of the number of participants ($N_{part}$) and
the number of binary nucleon-nucleon collisions ($N_{coll}$) in
158$\cdot A$~GeV Pb+Pb collisions over a wide impact parameter
range. A scaling of the transverse energy pseudorapidity density at
midrapidity as $\sim N_{part}^{1.08 \pm 0.06}$ and $\sim
N_{coll}^{0.83 \pm 0.05}$ is observed. For the charged particle
pseudorapidity density at midrapidity we find a scaling as $\sim
N_{part}^{1.07 \pm 0.04}$ and $\sim N_{coll}^{0.82 \pm 0.03}$. This
faster than linear scaling with $N_{part}$ indicates a violation of
the naive Wounded Nucleon Model.
%
% insert suggested PACS numbers in braces on next line
\PACS{
      {25.75.Dw}{Particle and resonance production}
%         \and
%       {PACS-key}{discribing text of that key}
     } % end of PACS codes
} %end of abstract

\maketitle
\begin{figure*}[bt]
   \centerline{\includegraphics{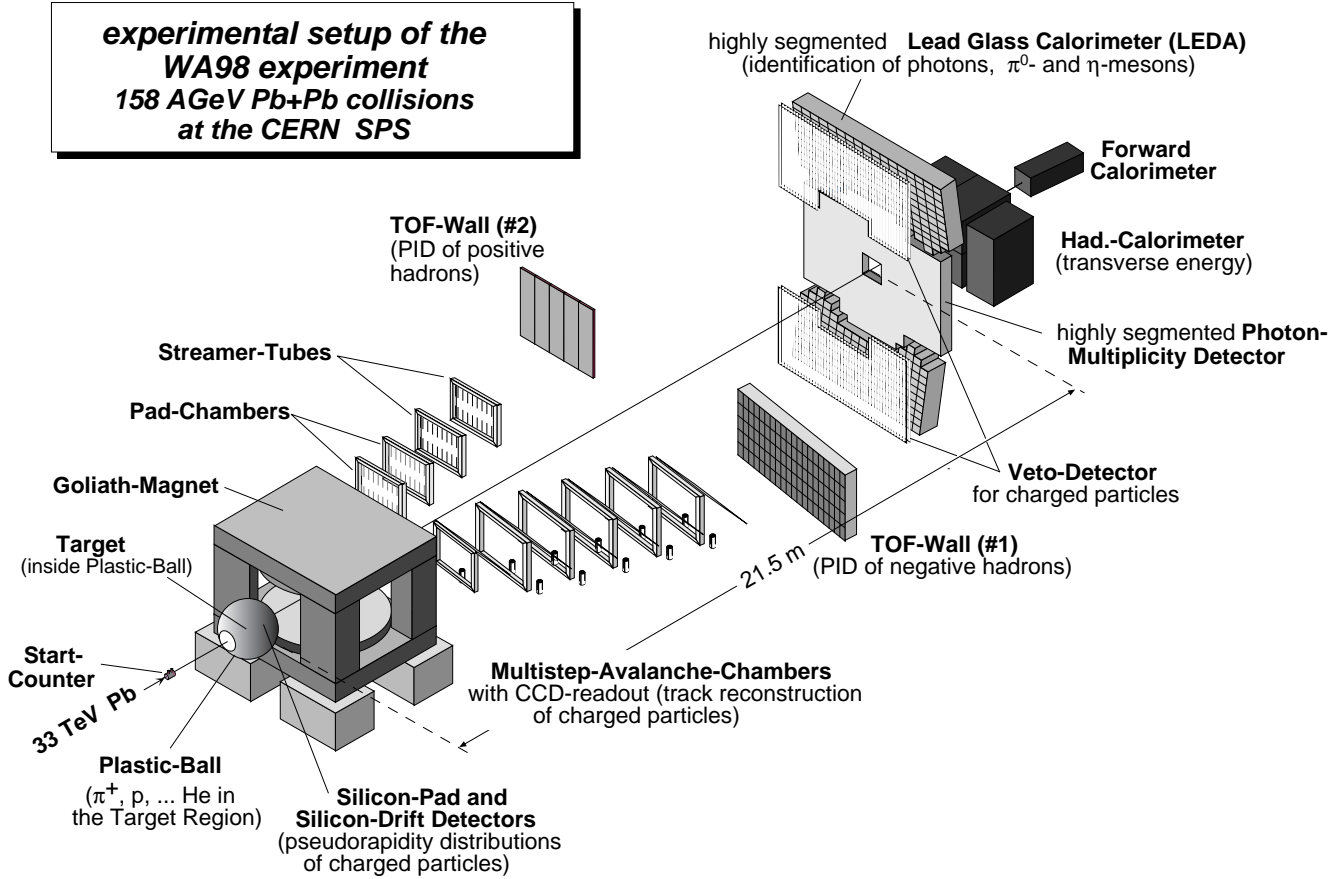}}
   \caption{The WA98 experimental setup.}
   \protect\label{fig:wa98}
\end{figure*}

\section{Introduction}
\label{sec:intro}

Heavy-ion collisions at ultrarelativistic energies probe nuclear matter at
high temperatures and densities. A major goal of these studies is the 
search
for a deconfined phase of nuclear matter. A necessary condition to 
reach such a phase transition is local equilibration as might be 
achievable through rescattering of the produced particles. Since the amount of 
rescattering should increase with the size of the reaction system, it 
is of interest to study these reactions over a wide range of 
centralities.

For hard processes, where cross sections are small, the 
naive expectation is a scaling of the particle yields with the 
number of binary collisions. Experimentally, the scaling of cross 
sections with target mass in p+A collisions was studied and the 
scaling was observed to be even stronger than this expectation 
\cite{cronin}. This was later attributed to multiple parton 
scattering in the initial 
state \cite{plb:krzywicki:cronin:79,zpc:lev:cronin:83}. From the same 
experiment it was also seen that particle production at 
intermediate $p_{T}$ shows a much weaker increase with target mass.

The gross features of particle production in nucleon-nucleus
collisions and reactions of light nuclei are well described in the framework 
of the Wounded Nucleon Model \cite{wnm}. 
In this model the transverse energy and particle production
in p+A and A+A reactions is calculated
by assuming a constant contribution from each participating nucleon.
This kind of scaling has also been observed by the WA80 collaboration in
reactions of $^{16}$O and $^{32}$S projectiles with various targets where
$dE_T/d\eta|_{max}$ was found to depend approximately linearly on 
the average total number of participants \cite{wa80_1}.

While a scaling with the number of collisions 
arises naturally in a picture of a superposition of nucleon-nucleon 
collisions, with a possible modification by initial state effects, the
Wounded Nucleon Model or participant scaling is more naturally related to 
a system with strong 
final state rescattering, where the incoming particles lose their 
memory and every participant contributes a similar amount of 
energy to particle production. The scaling behavior of particle 
production may therefore carry important information on the reaction 
dynamics. Various experimental signatures in heavy ion reactions 
require a comparison of observables for different system sizes. Therefore 
it is important to have a good understanding of these basic scaling 
properties. The scaling behavior can also be used as a valuable
test for models of particle production in heavy ion reactions (see 
e.g. \cite{vni}).

Furthermore, several observables in heavy ion reactions seem to 
show qualitative changes once a certain system size is reached.
Strangeness production is enhanced in S+S reactions compared 
to p+p, but seems to saturate for even larger nuclei 
(see e.g. \cite{qm97:sollfrank}). Recent results from 
the WA98 experiment \cite{wa98pi0} show a significant change of the shape 
of the $\pi^0$ $p_T$ spectrum in peripheral Pb+Pb collisions compared 
to p+p data. The shape, however, remains unchanged in the range of 
semi-central Pb+Pb collisions with about 50 participating nucleons up 
to central reactions. 

The NA50 collaboration has observed an anomalously suppressed $J/\psi$ 
yield in central Pb+Pb collisions in contrast to peripheral 
reactions \cite{na50} where the suppression of the $J/\psi$ yield can be 
explained by absorption in nuclear matter. 
This anomalous $J/\psi$ suppression provides an additional incentive to 
study the scaling behavior of particle production with the number of 
participants. Most models based on $J/\psi$ absorption by hadronic 
comovers assume the comover density to scale linearly with the number 
of participants and are then not able to fit the anomalous suppression 
\cite{jpsi1,jpsi2}. 
Only if the hadronic comover density scales substantially faster than 
linearly with the number of participants it is possible to obtain reasonable 
fits to the anomalous suppression.
It is therefore of interest to study in 
detail the centrality dependence of particle production and 
investigate its scaling properties with respect to the number of 
participants or collisions.

\section{Experiment and Data Analysis}
\label{sec:experiment}

The CERN experiment WA98 is a general-purpose apparatus which consists
of large acceptance photon and hadron spectrometers together with
several other large acceptance devices which allow to measure various
global variables on an event-by-event basis. The experiment took data
with the 158$\cdot A$~GeV $^{208}$Pb beams from the CERN SPS in 1994,
1995, and 1996.  The layout of the WA98 experiment as it existed
during the final WA98 run period in 1996 is shown in
Fig.~\ref{fig:wa98}. The data presented here were taken during the
1996 lead beamtime. The transverse energy and charged particle
distributions shown in this paper were measured with the magnetic
field of the Goliath magnet turned off. The minimum bias cross section
for this configuration was $\sigma_{mb} = (6260 \pm 280)$~mb. The
error of the minimum bias cross section relates to the uncertainty of
the target thickness and the uncertainty in the subtraction of the
contribution from interactions outside the target. The contribution
of these interactions was determined in special target-out runs. A
20\% uncertainty of this contribution was assumed in the calculation
of the uncertainty of the minimum bias cross section. The WA98
experiment took most of the data with the Goliath magnet turned on.
Since the analysis of the photon and neutral pion scaling requires
high statistics the respective spectra used here were measured in a
field-on configuration. The minimum bias cross section for this data
set was $\sigma_{mb} = (6440 \pm 300)$~mb.

The Zero Degree Calorimeter is located 30~m downstream of the target
and measures the total energy of all particles within an angle $\Theta
< 0.3 ^\circ$ relative to the beam axis in the laboratory system.  The
MIRAC calorimeter is placed 24~m downstream of the target
\cite{mirac}.  It consists of a hadronic and an electromagnetic
section and covers the pseudorapidity interval $3.5 < \eta <
5.5$. MIRAC plays the central role in the WA98 minimum bias trigger
where the measured $E_T$ is required to be above a minimum threshold.
The systematic errors of $dE_{T}/d\eta$ at midrapidity are dominated
by the correction for the differences in the response of the MIRAC to
hadronic and electromagnetic showers and to the extrapolation of the
distribution of $dE_{T}/d\eta$ to midrapidity. These combine to give
an overall systematic uncertainty of $\approx 20\%$ in the absolute
result for $dE_{T}/d\eta|_{max}$.  The centrality dependent part of
this uncertainty is much smaller and estimated to be approximately 
5\% only. The correction of $dE_{T}/d\eta$ due to interactions
outside the target for peripheral Pb+Pb reactions is less than 2\%. 

The charged particle multiplicity is measured with a circular Silicon
Pad Multiplicity Detector (SPMD) located 32.8~cm downstream of the
target \cite{spmd}. It consists of four quadrants each produced from
a 300~$\mu$m thick silicon wafer. This detector provides full
azimuthal coverage of the pseudorapidity region $2.35 < \eta < 3.75$
with 180 $\Phi$-bins and 22 $\eta$-bins. The pad size increases radially
to provide an approximately uniform pseudorapidity coverage. In
central Pb+Pb collisions the probability that a pad is hit by two or
more particles is not negligible. Therefore the multiplicity in an
$\eta$-ring is determined from the sum of the measured energy losses
of the charged particles traversing the $\eta$-ring divided by the
average energy loss per charged particle. The charged particle
pseudorapidity distribution is corrected for $\delta$-electrons
produced by lead ions traversing the 213 $\mu$m thick $^{208}$Pb
target foil. On average these electrons give rise to roughly 11
additional hits in the SPMD. This contribution has been determined
from beam triggers where no inelastic interaction took place. The
systematic error of $dN_{ch}/d\eta$ relates to the uncertainty in the
determination of the total energy loss of the charged particles in the
SPMD and to the correction for $\delta$-electrons. The uncertainty in
the energy loss measurement is estimated to result in a 3\% centrality
independent systematic error. The correction for $\delta$-electrons
at midrapidity ($dN_{\delta}/d\eta|_{mid} \approx 9$) is assumed to be
known with an accuracy of 10\% and contributes significantly to the
total uncertainty only for peripheral reactions. However, the
$dN_{ch}/d\eta$ distributions as shown in figure~\ref{fig:dndeta}
exhibit a slight asymmetry around midrapidity. We can force the
$dN_{ch}/d\eta$ distribution to be symmetric by arbitrarily increasing
the subtracted $\delta$-electron contribution. For central events the
$\delta$-electron yield has to be increased by roughly 80\%. This
factor decreases when going to semi-central and peripheral events.
It's difficult to imagine a physical reason for a much larger
$\delta$-electron production than what was measured in target-out
events. However, by making these extreme assumptions we estimate the
centrality dependent error of $dN_{ch}/d\eta$ by comparing the
$dN_{ch}/d\eta$ distributions with $\delta$-electrons subtracted
as measured in beam events with the $dN_{ch}/d\eta$ distributions that
were forced to be symmetric.
For peripheral and semi-central events this uncertainty typically is
of the order of $3-4$\% and decreases to 2\% in central events. The
correction of $dN_{ch}/d\eta$ measured with the SPMD in peripheral
events due to interactions outside the target is typically of the
order of 15\%. The uncertainty of $dN_{ch}/d\eta$ due to this
correction is estimated to be around 3\%.

The photon distributions used in this analysis were measured with the
LEDA spectrometer in the rapidity interval $2.3 < \eta < 3.0$.  This
detector is located 21.5~m from the target and consists of 10080
leadglass modules each read out by a photomultiplier. A streamer tube
array placed directly in front of LEDA was used as a charged particle
veto detector to correct for the charged hadron contamination in the
leadglass.  The remaining correction for neutrons and anti-neutrons
has been made based on simulation results using the GEANT
package~\cite{geant}.  The detection efficiency of photons in LEDA is
based on GEANT simulations and experimental data in order to take into
account the effects of overlapping showers which can result in a shift
in the measured transverse momentum.  These corrections require high
statistics and therefore only 8 centrality classes have been used for
the photon analysis here. These are the same 8 centrality classes used
in the analysis of the scaling of neutral pion production presented in
\cite{wa98pi0}.  The systematic error on the photon and neutral pion
multiplicities is estimated to be $\approx 10\% $ mainly originating
from corrections for efficiency and contamination.

\section{Model Calculations}
\label{sec:model_calc}

In the present analysis,  
the photon and charged particle scaling has been 
investigated with the centrality of the Pb+Pb collision
determined from the transverse energy, $E_T$, measured with the MIRAC
calorimeter. 
However, in the $E_T$ scaling analysis, the forward 
energy $E_{F}$ of projectile 
spectators measured with the Zero Degree Calorimeter has been used
for the centrality selection, in order to avoid auto-correlations.
Twenty-one centrality classes have been defined based on the measured $E_T$.
Each class corresponds to $5\%$ of the minimum bias cross section,
with an additional very central class corresponding to the 
$1\%$ most central events.
The ZDC cannot resolve the centrality of very peripheral collisions as
well as the MIRAC calorimeter and therefore the $E_T$ scaling analysis
is limited to centrality classes which correspond to 
more than approximately 40 participants.

\begin{figure}[tb]
   \centerline{\includegraphics[width=8.8cm]{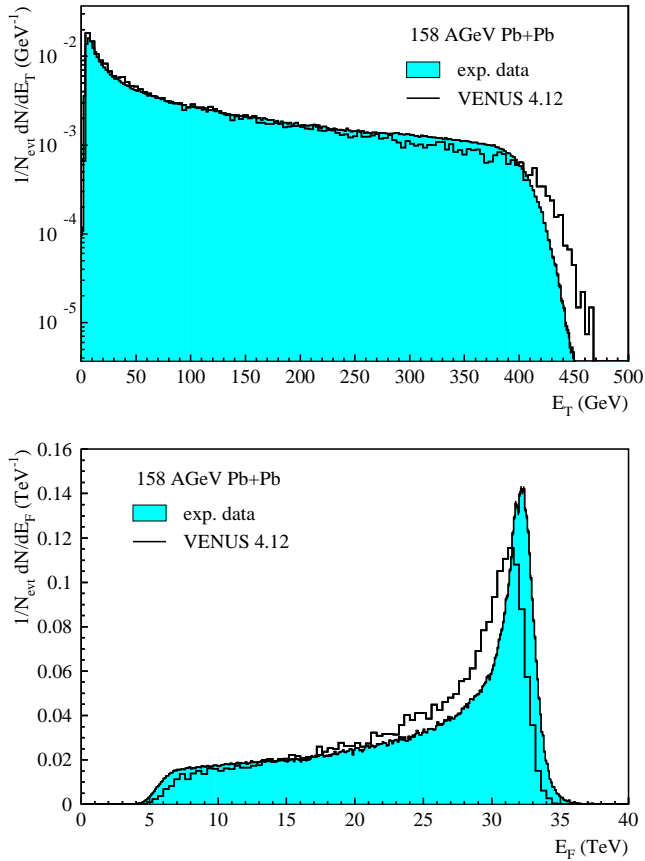}}
   \caption{Distributions of the transverse energy $E_{T}$ as measured 
   in MIRAC (upper graph) and the forward energy $E_{F}$ as measured with the 
   ZDC (lower graph) in 158$\cdot A$~GeV Pb+Pb collisions. Predictions of the 
   event generator VENUS 4.12 are included.}
   \protect\label{fig:etezdc}
\end{figure}

\begin{figure}[bt]
   \centerline{\includegraphics[width=8.8cm]{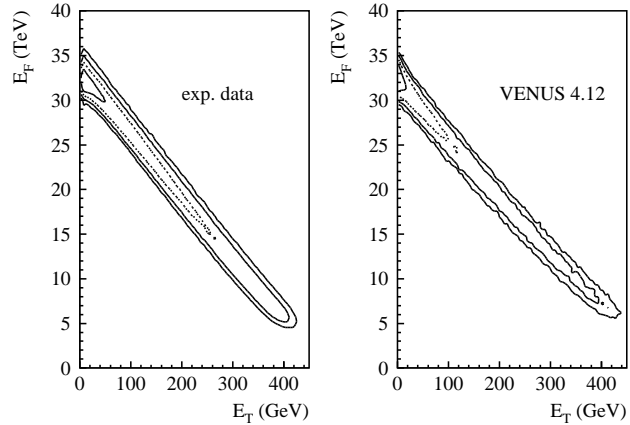}}
   \caption{Event-by-event distributions of the transverse energy $E_{T}$ 
   vs. the forward energy $E_{F}$ in 158$\cdot A$~GeV Pb+Pb collisions. The 
   left plot shows the experimental data and the right plot 
   predictions of the event generator VENUS 4.12.}
   \protect\label{fig:etvsezdc}
\end{figure}

The number of participants $N_{part}$ and collisions $N_{coll}$ for a
given centrality class have been determined from a simulation based on
the event generator VENUS 4.12 \cite{venus}. Control calculations were
made with the FRITIOF event generator \cite{fritiof}. The
definition of $N_{part}$ and $N_{coll}$ in these models is based on
a geometrical picture (Glauber model) where nucleons travel on
straight-line trajectories and the nucleon-nucleon cross section is
independent of the number of collisions a nucleon has undergone
before. The approach of using event generators allowed to take into
account the energy resolution of the ZDC and MIRAC
calorimeters. Furthermore, the minimum bias trigger efficiency has
been included in the simulation. Figure~\ref{fig:etezdc} shows a
comparison of the measured $E_{T}$ and $E_{F}$ distributions to the
VENUS simulations. The overall agreement in the $E_{T}$ distribution
between the data and the model is good.  However, the VENUS prediction
extends to slightly higher transverse energy for the most central
reactions. In the forward energy distribution, the strong peak for
peripheral reactions is not precisely reproduced, while the general
shape is quite similar.  Also the event-by-event anti-correlation of
$E_T$ and $E_{F}$ observed in the experimental data is in good
agreement with the model calculations (see figure~\ref{fig:etvsezdc}).
\begin{figure}[tb]
   \centerline{\includegraphics[width=8.8cm]{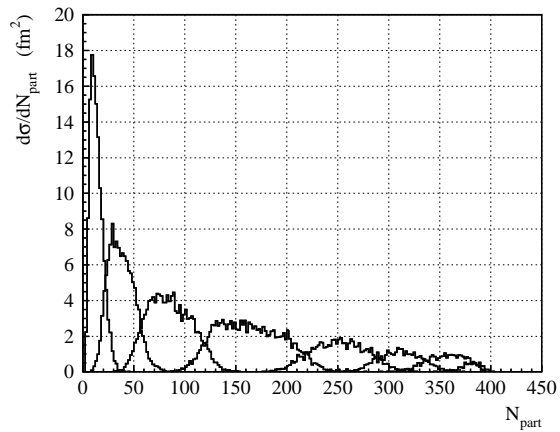}}
   \caption{Distributions of the number of participants $N_{part}$ 
   obtained from VENUS 4.12 for different centrality classes selected 
   by the transverse energy $E_{T}$.}
   \protect\label{fig:npartdist}
\end{figure}
\begin{figure}[bt]
   \centerline{\includegraphics[width=8.8cm]{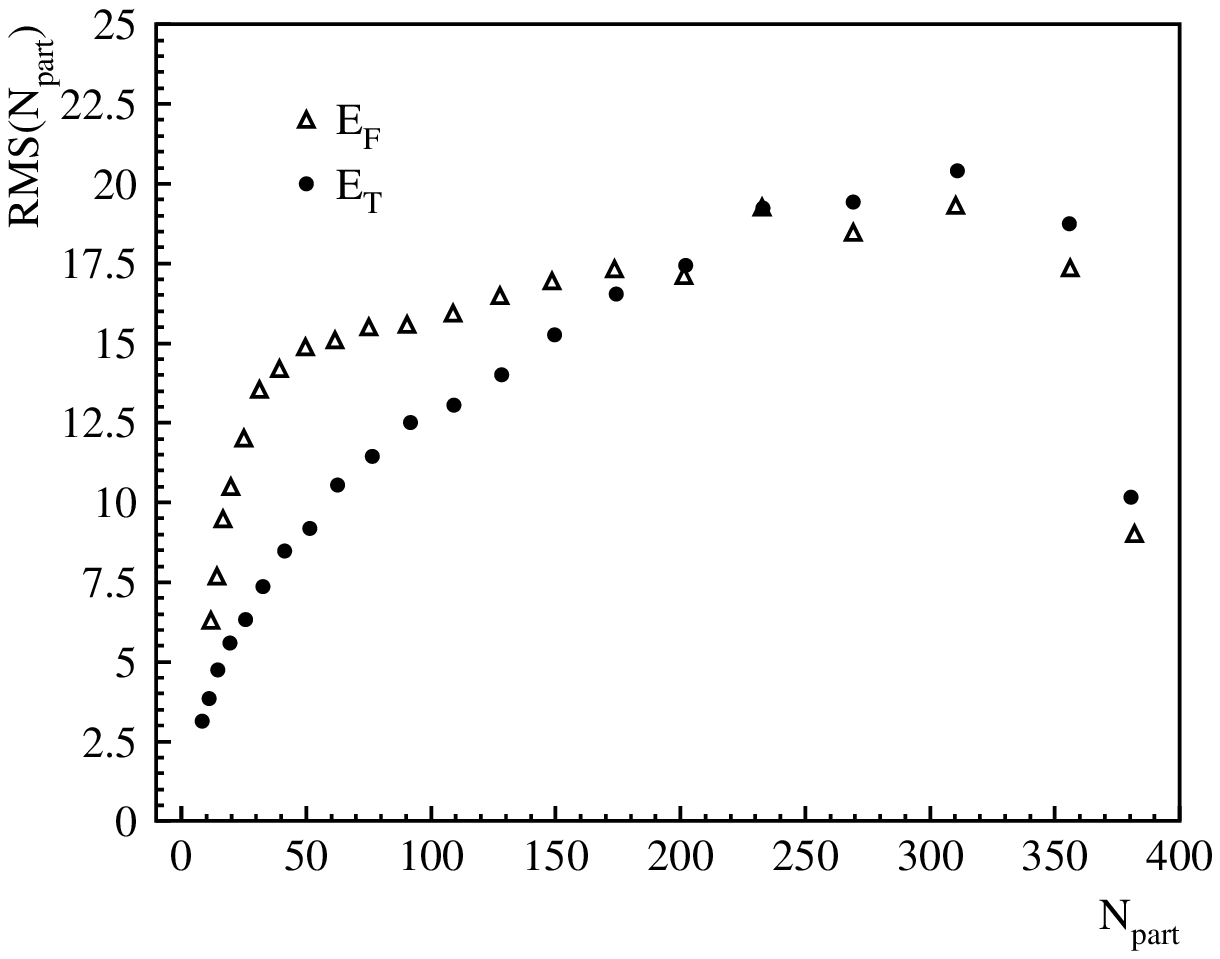}}
   \caption{RMS-values of the number of participants $N_{part}$ 
   obtained from VENUS 4.12 for different centrality classes selected 
   by the transverse energy $E_{T}$ and the forward energy $E_{F}$.}
   \protect\label{fig:npartrms}
\end{figure}

To obtain a robust estimate of 
$N_{part}$ and $N_{coll}$, which is less sensitive to discrepancies 
in the energy distributions, the centrality classes in the model have 
been chosen to represent the same absolute cross section as the data. 
The effect of the centrality cuts on the distributions of the 
number of participants can be seen from 
figure~\ref{fig:npartdist}, where distributions of $N_{part}$ for 
centrality classes corresponding to fractions of the minimum bias 
cross section of 0-1\%, 1-5\%, 5-10\%, 10-20\%, 20-40\%, 40-60\%, 
60-80\% and 80-100\% are shown. One can see that the limited 
acceptance, the detector resolution, and the fluctuations in particle 
production as implemented in the model lead to an overlap of the 
distributions of adjacent centrality classes. Nevertheless, it is observed 
that even a strong cut on the 1\% most central reactions yields 
a significantly different selection than e.g. the 5\% most central 
reactions. 

Figure~\ref{fig:npartrms} shows a summary of the RMS-values versus the 
average values of the distributions of $N_{part}$ for the 
centrality selections as used in the later analysis in this paper. 
While the classes selected with $E_{F}$ have a slightly smaller 
width for very central reactions, for peripheral reactions the 
resolution of the selection is much better using $E_{T}$.

The precise number of participants or number of collisions may, however, 
depend on the specific model assumptions. We have 
therefore performed a detailed study of the influence of these 
assumptions and other sources of systematic errors for the analysis 
presented here. A summary of these 
studies is given in appendix \ref{sec:app1}. 

\section{Results}
\label{sec:results}

\begin{figure}[tbh]
   \centerline{\includegraphics[width=8.8cm]{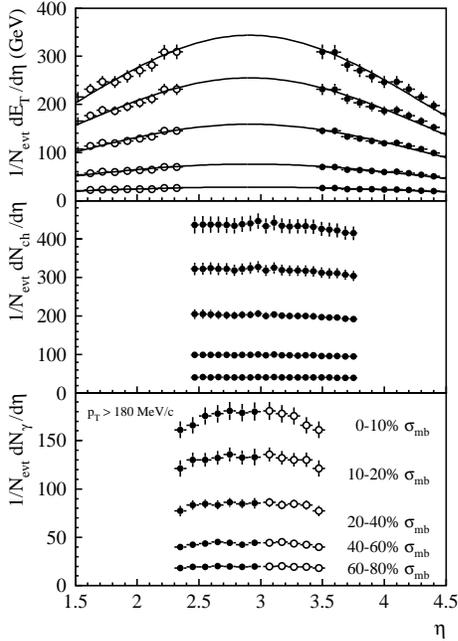}}
   \caption{Pseudorapidity distributions of transverse energy, 
            charged particles and photons measured in 
            158$\cdot A$~GeV Pb+Pb collisions of different centrality.
            Photons were measured above a lower transverse 
            momentum threshold of 180~MeV/$c$.
            The open symbols have been obtained by 
            reflecting the measured data points at midrapidity. }
   \protect\label{fig:dndeta}
\end{figure}

The pseudorapidity distributions for the transverse energy, the charged 
particle multiplicity, and the photon multiplicity are shown in 
figure~\ref{fig:dndeta} for five centrality classes.
All distributions have been corrected for possible contributions of 
reactions upstream and downstream of the target and in the target frame 
by subtracting the respective yield determined in target-out runs, see
\cite{reygers} for further details.
To obtain $dE_T/d\eta|_{max}$ the measured data points have 
been reflected at midrapidity ($\eta_{cm} = 2.91$) and fitted with a 
Gaussian.

\begin{figure}[tbh]
   \centerline{\includegraphics[width=8.8cm]{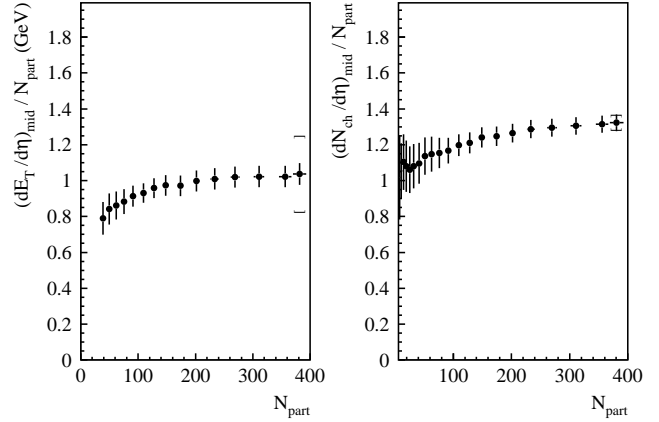}}
   \caption{Transverse energy and charged particle yields in 
            158$\cdot A$~GeV Pb+Pb reactions normalized to 
            the number of participants as a function of the number of
            participants. 
            For the number of participants an uncertainty as indicated
            in figure~\ref{fig:npart_check} was assumed. Furthermore,
            all centrality dependent errors of $dE_T/d\eta$ and
            $dN_{ch}/d\eta$ as described in section \ref{sec:experiment}
            were taken into account in the calculation of the error bars.
            The brackets indicate the centrality independent uncertainty
            of the overall $dE_T/d\eta$ and $dN_{ch}/d\eta$ scale. }
   \protect\label{fig:et_nch_per_part}
\end{figure}  
A first impression of the centrality dependence of $E_T$ and 
charged particle production can be obtained by normalizing the yields
to the number of participants. This is shown in 
figure~\ref{fig:et_nch_per_part}. Both for $E_T$ and the charged 
multiplicity the yield per participant increases when going
from peripheral to more central reactions. 
As described in the appendix we have also used the FRITIOF model
instead of VENUS to calculate the number of participants and
the number of nucleon-nucleon collisions. The results of these
two calculations for the number of participants are compared in
the appendix (fig.~\ref{fig:npart_check}) and used to obtain 
the estimated uncertainty of the number of participants
which was used in the calculation of the error bars in 
figure~\ref{fig:et_nch_per_part}. 

\begin{figure}[tb]
   \centerline{\includegraphics[width=8.8cm]{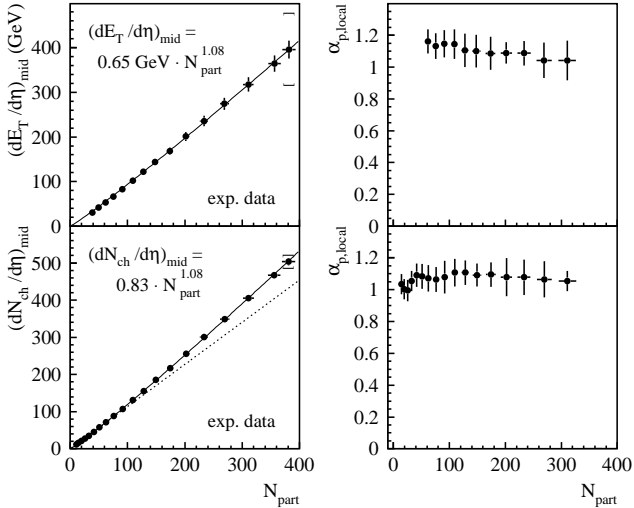}}
   \caption{Pseudorapidity density of $E_T$ and $N_{ch}$ at
            midrapidity as a function of the number of participants.
            The participant numbers shown here were calculated in a 
            VENUS simulation (denoted as calculation A in the appendix).
            Both for $E_T$ and $N_{ch}$ the fit results obtained from a 
            fit of all data points are shown.
            In order to demonstrate the stronger than linear increase
            of the data points a linear extrapolation of the charged 
            particle multiplicity in peripheral Pb+Pb reactions 
            ($N_{part} \approx 50$) is shown as a dotted line in the lower  
            left plot.
            The scaling behavior can be verified in more detail
            on the right panel where the local scaling exponents are shown.
            The local scaling exponents have been obtained from a fit
            of 5 neighboring $E_T$ and $N_{ch}$ data points. 
            }
   \protect\label{fig:etamax}
\end{figure}

The scaling behavior is studied in more detail in figure~\ref{fig:etamax}
which shows the dependence of the $E_T$ and $N_{ch}$ pseudorapidity 
densities at midrapidity on the number of participants.
The scaling behavior of these observables was parameterized as
\begin{equation}
  \left. \frac{dX}{d\eta}\right|_{mid} \sim N_{part}^{\alpha_p}
  ,\: N_{coll}^{\alpha_c},\quad X = E_T,\,N_{ch}\,. 
  \label{eq:scaling}
\end{equation}
This functional dependence gives a reasonable description of the data
for the entire centrality range. Taking the number of participants
from the VENUS calculation (denoted as calculation A in the appendix)
the charged particle scaling can be described by a scaling exponent
$\alpha_p = 1.08$. The calculations of the number of participants
using VENUS, or VENUS with an experimentally determined nucleon density
distribution, or FRITIOF (denoted as calculations A, B, and F in the
appendix) are all based on reasonable assumptions. These three calculations
give slightly different participant numbers. We quote the average of the
corresponding three scaling exponents $\alpha_p$ as the
final result. As described in the appendix the estimated uncertainty
of $\alpha_p$ due to the uncertainty of the number of participants is
0.036. In the fit from which we obtain the value of $\alpha_p$ 
we only take the statistical error of the data points into account. 
Since the statistical errors are small the fit error is negligible. 
In order to estimate the influence of the centrality dependent
errors of $dN_{ch}/d\eta$ on $\alpha_p$ we systematically move 
the data points within the error bars and repeat the fit.
From this procedure we estimate an uncertainty of 0.02 for $\alpha_p$
due to the centrality dependent errors of $dN_{ch}/d\eta$.
Adding all errors in quadrature we finally obtain 
$\alpha_p = 1.07 \pm 0.04$ for the charged particle scaling.   
   
For the $E_T$ scaling we perform a similar error analysis. An
additional uncertainty comes from the assumed centroid of the
$dE_T/d\eta$ distribution. Primarily due to massive particles like
protons and neutrons, the difference between pseudorapidity and
rapidity could lead to an increase of the centroid relative to
midrapidity ($y_{mid} = 2.91$). By varying the assumed $dE_T/d\eta$
centroid position in the $\eta$-range $2.91 \pm 0.3$ a corresponding
error of 0.02 was estimated for the $E_T$ scaling exponent $\alpha_p$.
Adding all errors in quadrature we obtain $\alpha_p = 1.08 \pm 0.06$.

In more detail, the relative scaling for different centralities can 
be judged from the \textit{local scaling exponent} $\alpha_{local}$ 
shown on the right hand side of figure~\ref{fig:etamax}.
These have been obtained from a fit of 5 neighboring $E_T$ and 
$N_{ch}$ data points. For the charged particles it can be seen
that the scaling remains approximately constant over the whole
centrality range. For $E_T$ the local scaling exponent appears to be 
almost constant in the range $N_{part} > 100$. However, below 
$N_{part} \approx 100$ the local scaling exponent seems to increase 
slightly. 

Considering the scaling with the number of binary nucleon-nucleon
collisions and averaging the results determined with calculations 
A, B, and F we obtain a similarly good description with
$\alpha_c = 0.82 \pm 0.03$ for the charged particle scaling.
For the $E_T$ scaling we obtain $\alpha_c = 0.83 \pm 0.05$.
These results are not surprising since for symmetric systems 
one naively expects a scaling relation to hold between the number of 
collisions and the number of participants:
\begin{equation}
    N_{coll} \propto N_{part}^{\alpha_{cp}} \,\,\, \mathrm{with} \,\,\, 
    \alpha_{cp} = \frac{4}{3}.
    \label{eq:collpart}
\end{equation}
Fits of the parameters extracted from VENUS simulations indeed yield a 
value of $\alpha_{cp} = 1.28$ which is close to the above value. For a 
scaling with $N_{part}^{1.08}$ this would lead to a behavior as 
$N_{coll}^{0.84}$.

\begin{figure}[tb]
   \centerline{\includegraphics[width=8.8cm]{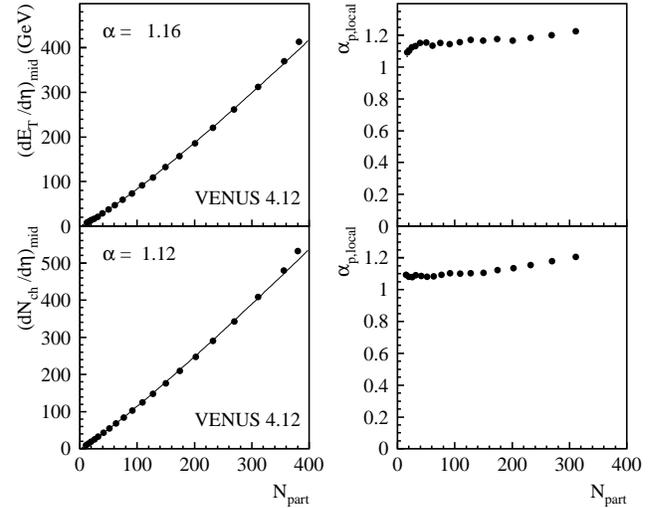}}
   \caption{Pseudorapidity density of $E_T$ and $N_{ch}$ at
            midrapidity as a function of the number of participants as 
            in figure~\ref{fig:etamax} from VENUS 4.12 simulations.
            }
   \protect\label{fig:etamaxv}
\end{figure} 
A similar analysis can be performed on the data obtained from the
VENUS simulation itself. The results of such an analysis are displayed
in figure~\ref{fig:etamaxv}. It is observed that the scaling exponents
are higher in the simulation, and that both for $E_T$ and $N_{ch}$ the
exponent $\alpha$ shows a tendency to increase with centrality.  We
note here that the extraction of the number of participants and the
number of binary nucleon-nucleon collisions is almost completely
independent of the scaling exponent present in the underlying event
generator. As described in section \ref{sec:model_calc} this is due to
the fact that the centrality classes defined in the model calculations
correspond to the same absolute cross section as the experimental
centrality classes.

\begin{figure}[tbh]
   \centerline{\includegraphics[width=8.8cm]{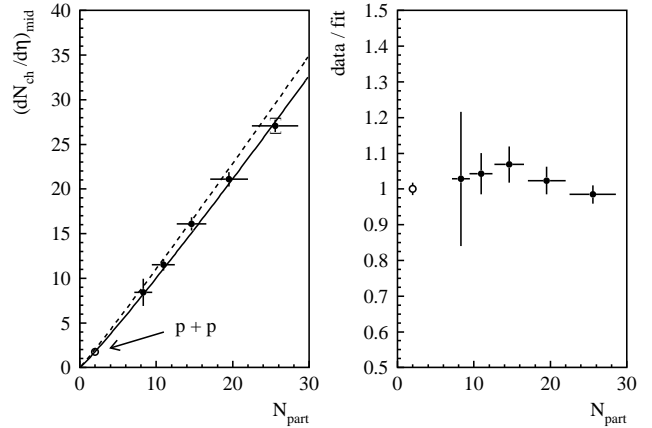}}
   \caption{Pseudorapidity density of $N_{ch}$ at
            midrapidity as a function of the number of participants 
            for p+p \cite{nch_pp} and Pb+Pb collisions.
            The number of participants shown here were calculated 
            using VENUS (calculation A in the appendix).
            The fit function plotted as a solid line is the 
            same as obtained in figure~\protect\ref{fig:etamax}. 
            The dashed function is the fit result using the participant
            values from the FRITIOF calculation (calculation F).
            On the right hand side the ratio of the data 
            (using participants from VENUS) to the fit function is shown.}
   \protect\label{fig:npp}
\end{figure}

It is interesting to extrapolate this scaling towards smaller system 
sizes and compare to the expectation from pp 
collisions \cite{nch_pp}. This has been done in 
figure~\ref{fig:npp}. It can be seen that the scaling obtained from 
charged particle production in Pb+Pb collisions extrapolates nicely 
to pp collisions.
In particular, there is no threshold effect visible in the charged particle
multiplicity when going to central Pb+Pb collisions.

\begin{figure}[tbh]
   \centerline{\includegraphics[width=8.8cm]{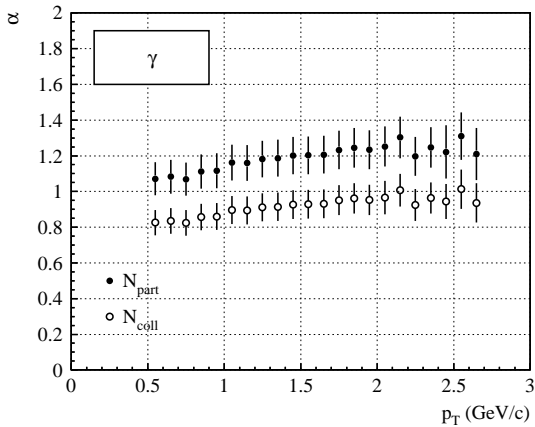}}
   \caption{Exponents for a scaling with the number of participants and
   nucleon-nucleon collisions for photons measured in LEDA as a 
   function of the transverse momentum. 
   Only reactions with $N_{part} \geq 30$ have been used.
   The data points shown were obtained using the number of participants
   from VENUS. The error bars reflect the uncertainty of the photon 
   measurement, the uncertainty in the calculation of $N_{part}$/$N_{coll}$ 
   and the fit error.
}
   \protect\label{fig:phot}
\end{figure}  
In a recent publication \cite{wa98pmd} we have discussed the 
systematics of inclusive photon production as measured with the WA98
Photon Multiplicity Detector. There it was found that 
the pseudorapidity density of photons at midrapidity scales with the 
number of participants as $N_{part}^{1.12 \pm 0.03}$. 
This is slightly larger, but consistent with the exponent from 
the present analysis of $E_T$ and $N_{ch}$. 
Photon production can be investigated in further detail with the photon 
spectrometer LEDA. It can very naturally be 
studied as a function of the transverse momentum. The 
corresponding scaling 
exponents $\alpha$ extracted are shown in figure~\ref{fig:phot}. 
The photon measurement in LEDA suffers from larger systematic 
uncertainties at low momenta, so only photons for $p_T > 500$~MeV/$c$ 
have been considered.
At a transverse momentum of $p_T \approx 500$ MeV/$c$ the inclusive 
photon yield shows a scaling behavior similar to that observed for
$E_T$ and $N_{ch}$. However, the extracted scaling exponents tend to rise 
with increasing $p_T$ and at $p_T \approx 2$~GeV/$c$ the scaling can 
be described as $\sim N_{part}^{1.2}$ and $\sim N_{coll}^{0.9}$.
 
\begin{figure}[tbh]
   \centerline{\includegraphics[width=8.8cm]{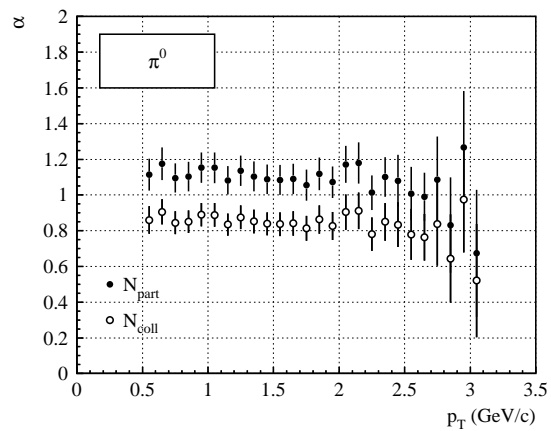}}
   \caption{Exponents for a scaling with the number of participants 
   and nucleon-nucleon collisions for neutral pions measured 
   in LEDA as a function of the transverse momentum. 
   Only reactions with $N_{part} \geq 30$ have been used.
   The data points shown were obtained using the number of participants
   from VENUS. The error bars reflect the uncertainty of the $\pi^0$ 
   measurement, the uncertainty in the calculation of $N_{part}$/$N_{coll}$ 
   and the fit error.}
   \protect\label{fig:pion}
\end{figure}

Since a large fraction of the inclusive photons originates from the decay 
of neutral pions it is of interest to compare the scaling of photons 
to that of neutral pions which have already been discussed in \cite{wa98pi0}.
These data have been reanalyzed \cite{wa98pi0_erratum} and the results of
the scaling exponents with respect to the number of participants are shown
in figure~\ref{fig:pion}.
The values are nearly constant at $\alpha \approx 1.1$ with a tendency 
to decrease towards higher transverse momenta. It should be noted that 
the extracted exponents are smaller compared to the values given in 
\cite{wa98pi0}. This is mostly due to a more sophisticated 
calculation of the number of participants used here.

We have recently published results on the production of direct photons
in 158$\cdot A$~GeV Pb+Pb collisions
\cite{wa98_direct_phot_lett,wa98_direct_phot}.  A significant yield of
direct photons at $p_T > 1.5$~GeV/$c$ is observed in central
collisions, while in peripheral collisions the photon production is
consistent with the yield expected from the decays of neutral pions,
$\eta$ mesons, and other hadrons. At $p_T \approx 2$~GeV/$c$ the
direct photon yield in central collisions amounts to roughly 20\% of
the photons from hadronic decays.  The scaling of the neutral pion
yield presented in figure~\ref{fig:pion} appears to be consistent with
the scaling of charged particles and the transverse energy, while the
photon yield at higher transverse momenta seems to increase more
strongly with the number of participants.  We assume that for a
relatively small yield of direct photons in central collisions, the
centrality dependence of the inclusive photons can still be described
with a scaling law as in equation~(\ref{eq:scaling}).  It's then
obvious that the production of direct photons in central Pb+Pb
collisions necessarily increases the scaling exponent $\alpha$ of the
inclusive photons at high transverse momenta.  In this respect the
behavior of the extracted scaling exponents for inclusive photon and
neutral pion production are consistent with the direct photon excess
observed in central Pb+Pb collisions.  However, due to the
uncertainties of the scaling exponents for photons and neutral pions
it is not possible to draw quantitative conclusions about direct
photon production from figures~\ref{fig:phot} and \ref{fig:pion}.

In simple multiple collision models a heavy ion reaction is regarded as a
sequence of independent nucleon-nucleon collisions which can be described 
as in free space~\cite{wong_1,wong_2}. 
After a projectile nucleon suffers an inelastic collision
the assumption of local baryon number conservation assures that a 
baryon-like object is still present. This baryon-like object is assumed to
contribute to the particle production in subsequent collisions with the
same cross section as the initial nucleon.
In this picture the contribution of each nucleon-nucleon collision is 
added
incoherently which leads to a linear scaling of $E_T$ and particle 
production 
with the number of binary collisions.
If the energy degradation in each nucleon-nucleon reaction is taken into 
account a reasonable description of $E_T$ and particle production can
be obtained. The approximate scaling as $N_{coll}^{0.83}$ for 
the transverse energy and charged particles may be used to obtain 
information on the average energy degradation in a nucleon-nucleon collision.
\begin{figure}[hbt]
   \centerline{\includegraphics[width=8.8cm]{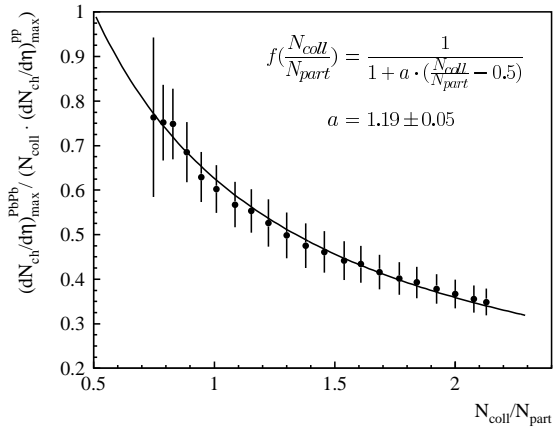}}
   \caption{Pseudorapidity density of $N_{ch}$ at midrapidity in Pb+Pb 
   collisions normalized to the number of collisions and the charged 
   particle density in pp collisions. The solid line shows a fit with 
   equation~(\ref{eq:wong2}).}
   \protect\label{fig:nchncoll}
\end{figure}  

A simple way of investigating this hypothesis is to study the particle 
production per binary collision as a function of the effective 
thickness $x$ of the two nuclei, which might be characterized in the 
following way~\cite{wong_1,wong_2,wong_book}:
\begin{equation}
    \frac{dN_{ch}}{d\eta}(AA) = N_{coll} \cdot f(x) \cdot 
    \frac{dN_{ch}}{d\eta}(pp).
    \label{eq:wong1}
\end{equation}
Here $f(x)$ should describe the effect of energy degradation on particle 
production with $x$ being a suitable thickness variable -- we have chosen 
$x \equiv N_{coll}/N_{part}$. Figure~\ref{fig:nchncoll} shows the 
pseudorapidity density of $N_{ch}$ at midrapidity in Pb+Pb collisions 
normalized to the number of collisions and the charged particle density 
in pp collisions. The data show a continuous decrease of the 
multiplicity per collision for increasing $x$, i.e. the more 
collisions a participant suffers, the smaller is the contribution of 
each collision to particle production. 
As an example, this can be illustrated more quantitatively for the 
centrality classes with $N_{coll}/N_{part} \approx 1$, i.e. for 
the case that each participating nucleon on the average suffers 
two nucleon-nucleon collisions. For these reactions the actual 
charged particle density at midrapidity is 40\% lower 
than one would expect from a linear scaling of the charged particle
density in pp reactions with the number of binary nucleon-nucleon collisions. 
In a simple multiple collision picture this means that on the average 
the second collisions of each participant contributes only 20\% of the 
yield of the first collision to the charged multiplicity at midrapidity.

In proton-nucleus reactions multiplication of the charged particle 
multiplicity observed in pp reactions by the number of binary
nucleon-nucleon collisions overestimates the measured multiplicity 
by $20-30\%$ \cite{pA_data,wong_3}. 
In case of proton-nucleus reactions the target participants suffer exactly
one nucleon-nucleon collision. This is of course not true in AA collisions.
Nevertheless, it is interesting to apply the same recipe to heavy ion
reactions. Figure~\ref{fig:nchncoll} shows that multiplying the pp yield
with the number of binary nucleon-nucleon collisions gives as much 
as 60\% too many charged particles in central Pb+Pb reactions.

We have attempted to fit the data in figure~\ref{fig:nchncoll} with
the form:
\begin{equation}
    f_{fit}\left( \frac{N_{coll}}{N_{part}} \right) =
    \left[ 1 + a \left( \frac{N_{coll}}{N_{part}} - 0.5 \right) 
    \right]^{-1}.
    \label{eq:wong2}
\end{equation}
This function basically represents a first order Taylor approximation
of the function $1/f$ in equation~(\ref{eq:wong1}). 
A good fit is obtained with $a = 1.19 \pm 0.05$.

In the case of the Wounded Nucleon Model the eventual fragmentation of
an excited nucleon after an inelastic collision is not affected by further
collisions with other nucleons, no matter how many times it is successively
struck. The observation of a stronger than linear increase 
of $E_T$ and $N_{ch}$ with the number of participants indicates
that this model is only approximately correct.
It can be clearly seen, regarding the two possible centrality 
variables investigated here, that the number of participants is better 
suited, because the scaling exponent is closer to one compared to the number 
of collisions. 

\begin{figure}[hbt]
   \centerline{\includegraphics[width=8.8cm]{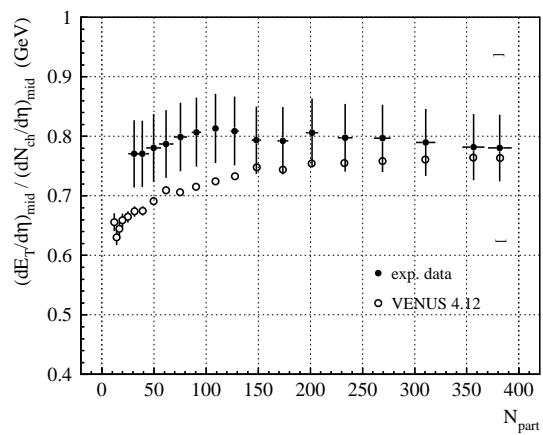}}
   \caption{$\langle E_T \rangle /\langle N_{ch} \rangle|_{mid}$ 
            as a function of the number of participants. 
            For comparison results of VENUS 4.12 calculations are included. 
            The error bars indicate the centrality dependent errors of 
            the experimental result.
            The overall uncertainty of $\langle E_T \rangle /\langle 
            N_{ch} \rangle|_{mid}$ which is mainly due to the uncertainty
            in $(dE_T/d\eta)_{mid}$ is indicated by brackets.}
   \protect\label{fig:etperch}
\end{figure}

With the present data it is possible to determine the average transverse 
energy per charged particle at midrapidity
 \begin{equation}
     \langle E_T \rangle /\langle N_{ch} \rangle|_{mid} \equiv 
     \langle dE_T/d\eta|_{mid} \rangle / \langle dN_{ch}/d\eta|_{mid} \rangle,
     \label{eq:etpnch}
\end{equation}
a quantity that can be seen as a measure of the global mean transverse 
momentum averaged over all particle species. 
$\langle E_T \rangle /\langle N_{ch} \rangle|_{mid}$ is plotted in  
figure~\ref{fig:etperch} as a function of centrality, 
represented by the number of participants.
For this figure both $dE_T/d\eta$ and $dN_{ch}/d\eta$ were evaluated in 
identical centrality classes, defined with the forward energy $E_F$.

$\langle E_T \rangle /\langle N_{ch} \rangle|_{mid}$ appears to
increase up to a system size of $N_{part} \approx 100$ which
corresponds to an impact parameter of $b \approx 9$~fm. For more
central collisions $\langle E_T \rangle /\langle N_{ch}
\rangle|_{mid}$ levels off at a value of 0.80~GeV. This value is
slightly higher than the maximum $\langle E_T \rangle /\langle N_{ch}
\rangle|_{mid} \approx 0.66$~GeV observed in 200$\cdot A$~GeV S+Au and
S+Al reactions~\cite{wa80_2}. VENUS 4.12 predicts a qualitatively
similar behavior for $\langle E_T \rangle /\langle N_{ch}
\rangle|_{mid}$, while the absolute value is approximately 100~MeV
lower at $N_{part} \approx 100$. 
One may also note that the VENUS results continue to rise by
$\approx 50 \, \mathrm{MeV}$ when going from $N_{part} \approx 100$ to
$N_{part} \approx 400$, while the experimental data appear to be
completely flat in this region.  A similar saturation with increasing
number of participants as observed here has been seen in the
(truncated) mean $p_T$ of neutral pions with $p_T > 400$~MeV/$c$
produced in Pb+Pb reactions \cite{wa98pi0}.  A natural explanation of
such a behavior would be the assumption that thermalization is reached
once the system exceeds a certain minimum size.

\section{Conclusions}
\label{sec:conclusions}

\begin{sloppypar}
We have analyzed the dependence of transverse energy and charged
particle pseudorapidity distributions and photon transverse momentum
spectra in 158$\cdot A$~GeV Pb+Pb collisions on the number of
participants and the number of binary nucleon-nucleon collisions. A
scaling behavior as $N_{part}^{1.07 \pm 0.04}$ and $N_{coll}^{0.82 \pm
0.03}$ describes the charged particle production over the whole impact
parameter range.  The $E_T$ production was studied for collisions with
more than approximately 40 participants. In this centrality range the
$E_T$ production scales as $N_{part}^{1.08 \pm 0.06}$ and
$N_{coll}^{0.83 \pm 0.05}$.

Photons at $p_T \approx 500$~MeV/$c$ show a scaling behavior similar to
the scaling of $E_T$ and $N_{ch}$.
The $p_T$-dependence of the photon scaling was studied and a
rise of the extracted scaling exponents with increasing transverse
momentum was found.

We have studied the transverse energy per charged particle as a 
function of centrality and found an indication of an increase from 
peripheral to semi-central collisions with approximately 100 participants 
with a subsequent saturation for larger systems.

While the global variables like $E_{T}$ and charged particle 
multiplicity seem to scale closer to the number of participants 
than to the number of binary collisions, there is a clear participant 
scaling violation compared to a purely linear dependence. This scaling 
violation might e.g. have consequences for the suppression of 
$J/\psi$ production from comovers, since the implied central particle 
densities are considerably larger than estimated based on a linear 
scaling.
\end{sloppypar}

\begin{appendix}

\section{Systematic uncertainties in model 
         calculations \protect\label{sec:app1}}

In order to obtain an estimate of the systematic uncertainties in the
calculation of the number of participants and the number of
nucleon-nucleon collisions we have varied several assumptions in the
model calculations. The dependence of the particle and transverse
energy yield on the number of participants and collisions is described
with the scaling exponents $\alpha_p$ and $\alpha_c$ in this paper.
In this section we investigate how the extracted scaling exponents
$\alpha_p$ and $\alpha_c$ for the charged particle yield (with similar
conclusions for the other observables) are affected by the different
model assumptions.

In the VENUS 4.12 simulations used to obtain the number of participants and 
collisions we have varied 
\begin{itemize}
  \item the parameterization of the nucleon density distribution,
  \item the energy resolution of MIRAC and
  \item the minimum bias cross section. 
\end{itemize}  
As a cross check we have also calculated the number of participants and 
collisions using the event generator FRITIOF \cite{fritiof}.

The nuclear density profile used in VENUS and FRITIOF is an effective
parameterization using a Woods-Saxon shape:
\begin{equation}
    \rho(r) = \rho_{0} \cdot \frac{1}
    {1 + \exp\left(\frac{r-R}{a} \right)}.
    \label{eq:woods-saxon}
\end{equation}
However, the two models make slightly different assumptions for the nuclear 
radius $R$ and the diffuseness parameter $a$.
The VENUS parameterization is  
\begin{equation}
  \label{eq:ra_venus}
  \begin{array}{l@{\:=\:}l}
  R_{\mathrm{VEN}} & (1.19 A^{1/3} - 1.61 A^{-1/3}) \; \mathrm{fm} \\
  a_{\mathrm{VEN}} & 0.54 \; \mathrm{fm} \\
  \end{array}
\end{equation}
which results in $R_{\mathrm{VEN}} \approx 6.78 \, \mathrm{fm}$ for a
lead nucleus.
The radius parameter in FRITIOF for nuclei with $A > 16$ is calculated as
\begin{equation}
  \label{eq:r_fritiof}
  \begin{array}{l@{\:=\:}l}
  R_{\mathrm{FRI}} & r_0 \cdot A^{1/3} \quad \mathrm{with}           \\
  r_0              & 1.16 \cdot (1-1.16 A^{-2/3}) \; \mathrm{fm}.    \\ 
  \end{array}
\end{equation}
This gives $R_{\mathrm{FRI}} \approx 6.65  \, \mathrm{fm}$ for lead nuclei.
The diffuseness parameter $a$ is taken to be slightly $A$-dependent in 
FRITIOF and lies in the range $0.47 \, \mathrm{fm} - 0.55 \, \mathrm{fm}$. 
For lead nuclei FRITIOF uses 
\begin{equation}
  \label{eq:a_fritiof}
  a_{\mathrm{FRI}} = 0.545  \, \mathrm{fm}.
\end{equation}

\begin{figure}[hbt]
   \centerline{\includegraphics[width=8.8cm]{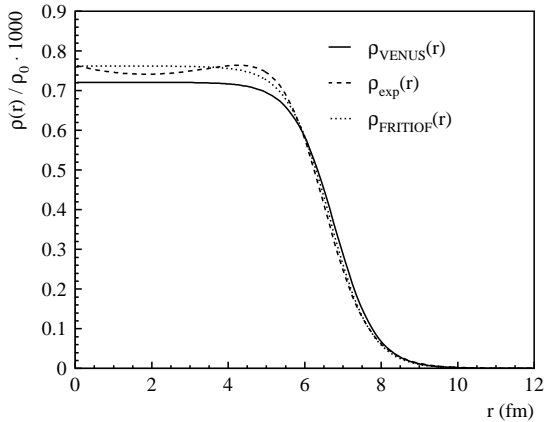}}
   \caption{Different nuclear density distributions used in the 
   calculations of the number of participants and the number of 
   collisions. The solid line shows the distribution implemented in 
   the VENUS 4.12 Monte-Carlo model (equation (\protect\ref{eq:woods-saxon})
   with the parameters (\protect\ref{eq:ra_venus})), 
   the dashed line shows a 
   parameterization of the charge distribution obtained from electron 
   scattering (equation (\protect\ref{eq:rhoexp})) and the dotted 
   line shows the density used in the 
   FRITIOF model (equation (\protect\ref{eq:woods-saxon}) with the parameters
   (\protect\ref{eq:r_fritiof}) and (\protect\ref{eq:a_fritiof})).}
   \protect\label{fig:nudens}
\end{figure}

Electron scattering experiments have shown, however, that the density 
distribution has a slightly more complicated 
structure \footnote{Strictly this applies only to the charge 
distribution. The true nucleon distribution, especially in the inner 
part of the nuclei, is not experimentally accessible.}. For a 
comparison we will use a parameterization fitted to electron scattering 
data on $^{208}\mathrm{Pb}$ presented in \cite{prl:frois}:
\begin{equation}
    \rho_{\mathrm{exp}}(r) = \rho_{0} \cdot \frac{c_1 + c_2 r + c_3 r^2}
    {1 + \exp\left(\frac{r-R_{\mathrm{exp}}}{a_{\mathrm{exp}}}\right)}
    \label{eq:rhoexp}
\end{equation}
with
\begin{displaymath}
    c_1 = 0.0633, \, c_2 = -0.002045 \, \mathrm{fm}^{-1}, 
    c_3 = 0.000566 \, \mathrm{fm}^{-2}
\end{displaymath}
and
\begin{displaymath}
    R_{\mathrm{exp}} = 6.413 \, \mathrm{fm}, \,\,  
    a_{\mathrm{exp}} = 0.5831 \, \mathrm{fm}.
\end{displaymath}
The three different distributions are shown in figure 
\ref{fig:nudens}. It can be seen that the overall agreement is 
quite good. The default distribution of VENUS 
has a slightly larger radius, while the parameterization of the 
experimental data shows small oscillations compared to the other two 
distributions.

With respect to the experimental resolution we have varied the energy 
resolution of the MIRAC calorimeter in the simulations. The measured  
values of the resolution \cite{mirac} are for the electromagnetic section:
\begin{equation}
    \frac{\sigma_{\mathrm{em}}}{E} = \frac{17.9 \, 
    \%}{\sqrt{E/\mathrm{GeV}}} 
\end{equation}
and for the hadronic section:
\begin{equation}
    \frac{\sigma_{\mathrm{had}}}{E} = \frac{46.1 \, 
    \%}{\sqrt{E/\mathrm{GeV}}}.
\end{equation}
This has been arbitrarily worsened to 
\begin{equation}
    \frac{\sigma_{\mathrm{em,had}}}{E} = \frac{85 \, 
    \%}{\sqrt{E/\mathrm{GeV}}} 
\end{equation}
for both sections of the calorimeter.

For an accurate determination of the number of participants and the
number of nucleon-nucleon collisions it is necessary that the
experimental minimum bias threshold is reproduced in the simulation.
In section~\ref{sec:experiment} we stated the two sources in the
determination of the experimental minimum bias cross-section: the
error of the target thickness and the error due to subtraction of
interactions outside the target. In our approach of calculating the
number of participants and nucleon-nucleon collisions we define 
centrality classes based on the simulated $E_T$ (in case of
the $N_{ch}$ scaling analysis) which correspond to the same absolute
cross-sections as the respective classes for the measured $E_T$.
The uncertainty of the minimum bias cross-section due to the
error of the target thickness therefore directly leads to an 
error of the $N_{part}$-values which affects the entire centrality range.
However, the uncertainty in the correction for interactions outside
the target affects the experimental $d\sigma/dE_T$ distribution
only in the peripheral range. More precisely, since the maximum $E_T$
measured in target-out events is around 75 GeV, according to 
table~\ref{tab:npartcoll} only the range $N_{part} < 50$ is affected.

The measured target thickness is $213 \pm 3 \, \mathrm{\mu m}$. The
uncertainty of the target thickness contributes a relative error of
roughly 1.5\% to the error of the experimental minimum bias cross
section. In order to check the influence on the extracted scaling
exponent we arbitrarily increase the minimum bias threshold in the
simulation by roughly 2\% to $\sigma_{mb}^{sim} = 6386 \,
\mathrm{mb}$.  We have furthermore checked the error of $\alpha_p$ due
to the uncertainty of the minimum bias cross section that relates to
the subtraction of non-target contributions. To do this, we have used an 
$E_T$ distribution without correction for reactions outside the 
target to translate the experimental $E_T$ cuts into cross section
cuts. The apparent minimum bias cross section for this $E_T$ distribution 
was $\sigma_{mb} = 6530$~mb.  

The different calculations of the number of participants and collisions
are summarized in the following list:
\begin{enumerate}
    \item[A] \protect\label{caseA} 
    VENUS 4.12 calculations using the standard settings for 
    the density distribution, the experimental resolution and the
    minimum bias cross section.

    \item[B] \protect\label{caseB}
    VENUS 4.12 calculations as in case \ref{caseA} with a modified 
    density distribution according to equation (\protect\ref{eq:rhoexp}).

    \item[C] \protect\label{caseC} 
    VENUS 4.12 calculations as in case \ref{caseA} with a 
    modified MIRAC resolution.
   
    \item[D] \protect\label{caseD} 
    VENUS 4.12 calculations as in case \ref{caseA} with an a minimum bias
    cross section increased to $\sigma_{mb}^{sim} =  6386 \, \mathrm{mb}$.

    \item[E] \protect\label{caseE}
    VENUS 4.12 calculations with cross section cuts derived from an 
    experimental $E_T$-spectrum that was not corrected for interactions
    outside the target (apparent minimum bias cross section: 
    $\sigma_{mb} = 6530$~mb).
  
    \item[F] \protect\label{caseF}
    FRITIOF calculations using standard settings as in calculation A.
\end{enumerate}

\begin{figure}[hbt]
   \centerline{\includegraphics[width=8.8cm]{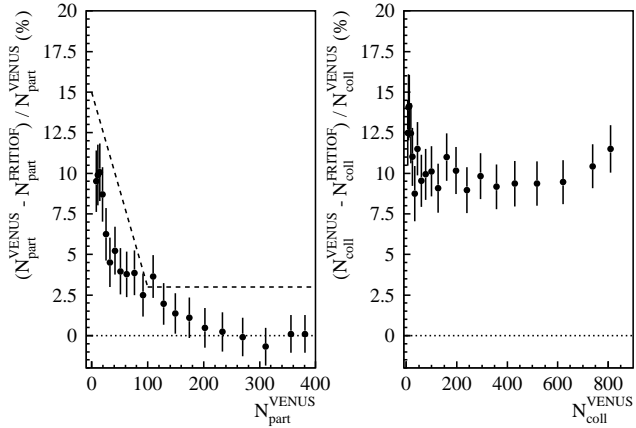}}
   \caption{Relative difference of the number of participants (left) and
     the number of nucleon-nucleon collisions (right) calculated with VENUS 
     (calculation A) and FRITIOF (calculation F).
     The dashed line on the left plot indicates the assumed uncertainty 
     of the participant scale that is used in the calculation of
     the error bars in figure~\ref{fig:et_nch_per_part}.} 
   \protect\label{fig:npart_check}
\end{figure}
As an example, the number of participants and collisions from calculation
A and F are compared in figure \ref{fig:npart_check}.
In peripheral Pb+Pb reactions VENUS gives up to 10\% more participants
than FRITIOF whereas in central reactions both simulations yield almost
identical results. Almost independent of centrality the number of collisions
from VENUS is roughly 10\% higher than the FRITIOF result.
The results of calculation A and F together 
with the experimental $E_T$ intervals are given in table \ref{tab:npartcoll}.
\begin{table}[bht]
\begin{center}
\begin{tabular}{|r|r|r|r|r|r|}
\hline
  \multicolumn{1}{|c|}{\%} & $E_{T}^{min}$    &   
  $N_{part}$ & $N_{coll}$ & $N_{part}$ & $N_{coll}$   \\
  \multicolumn{1}{|c|}{of c.s.} & (GeV) & 
  \multicolumn{2}{c|}{VENUS} & \multicolumn{2}{c|}{FRITIOF}  \\
  \hline \hline
     1 & 398.8 & 380.7 & 810.7 & 380.3 & 717.4 \\
     5 & 355.8 & 355.8 & 739.4 & 355.4 & 662.3 \\
    10 & 313.1 & 310.9 & 621.7 & 313.0 & 562.9 \\
    15 & 275.2 & 269.7 & 518.5 & 270.0 & 469.9 \\
    20 & 239.8 & 233.3 & 429.5 & 232.8 & 389.4 \\
    25 & 208.0 & 202.0 & 357.2 & 201.0 & 324.5 \\
    30 & 179.2 & 174.2 & 293.6 & 172.3 & 264.7 \\
    35 & 153.7 & 149.3 & 240.1 & 147.3 & 218.6 \\
    40 & 130.3 & 128.2 & 197.3 & 125.7 & 177.2 \\
    45 & 109.7 & 109.3 & 159.4 & 105.4 & 141.8 \\
    50 &  91.2 &  91.4 & 126.1 &  89.2 & 114.6 \\
    55 &  74.8 &  76.2 &  99.2 &  73.3 &  89.2 \\
    60 &  60.4 &  62.6 &  76.7 &  60.3 &  69.1 \\
    65 &  47.9 &  51.2 &  59.1 &  49.2 &  53.4 \\
    70 &  37.0 &  41.3 &  44.8 &  39.1 &  39.6 \\
    75 &  27.9 &  32.4 &  32.6 &  30.9 &  29.8 \\
    80 &  20.5 &  25.5 &  24.2 &  23.9 &  21.5 \\
    85 &  14.7 &  19.5 &  17.3 &  17.8 &  15.1 \\
    90 &  10.3 &  14.6 &  12.1 &  13.1 &  10.4 \\
    95 &   6.9 &  10.9 &   8.6 &   9.8 &   7.4 \\
   100 &   0.0 &   8.3 &   6.2 &   7.5 &   5.4 \\
\hline
\end{tabular}
\end{center}
\caption{The number of participants and binary collisions for 
different centrality classes obtained with the measured transverse 
energy in Pb+Pb collisions calculated from VENUS and FRITIOF (see text).}
\protect\label{tab:npartcoll}
\end{table}

For the case of the $N_{ch}$ scaling the impact of the different model
assumptions on the extracted scaling exponents $\alpha$ is summarized
in table \ref{tab:alpha_check}.  Considering the scaling with the
number of participants we take the average of the $\alpha_p$ values
from calculations A, B and F as our final result since all three
calculations are based on reasonable assumptions. The maximum
difference of 0.03 between the $\alpha_p$ values from these
calculations is taken as one contribution to the systematic error.  By
adding the deviations of the results from calculations C, D and E from
the mean value 1.07 in quadrature we estimate the total
systematic error of $\alpha_p$ related to the uncertainty of the
number of participants to be 0.036. The same prescription yields an
uncertainty of 0.024 for the exponent $\alpha_c$ that described the
scaling with the number of binary nucleon-nucleon collisions.
\begin{table}
\begin{center}
\begin{tabular}{|c|r|r|r|r|}
\hline
  calculation                            & 
  \multicolumn{1}{|c|}{$c_p$}            & 
  \multicolumn{1}{|c|}{$\alpha_p$}       & 
  \multicolumn{1}{|c|}{$c_c$}            &
  \multicolumn{1}{|c|}{$\alpha_c$}       \\
\hline \hline
 A  & 0.83 & 1.08 & 1.94 & 0.83 \\
 B  & 0.88 & 1.07 & 2.15 & 0.81 \\
 C  & 0.86 & 1.07 & 1.99 & 0.83 \\
 D  & 0.98 & 1.05 & 2.13 & 0.83 \\ 
 E  & 0.87 & 1.07 & 2.01 & 0.82 \\
 F  & 0.97 & 1.05 & 2.13 & 0.83 \\
\hline 
\end{tabular}
\end{center}
\caption{Influence of different model assumptions on the extracted exponents 
  $\alpha_p$ and $\alpha_c$ which describe the scaling of
  the charged particle yield with $N_{part}$ and $N_{coll}$ according
  to equation \ref{eq:scaling}. In addition to the scaling exponents we 
  quote the proportionality constants for the scaling with $N_{part}$ 
  ($c_p$) and $N_{coll}$ ($c_c$) for each calculation.
  The full centrality range was used in the fit of 
  equation~(\ref{eq:scaling}) to the measured charged particle
  yields.} 
\label{tab:alpha_check}
\end{table}

\end{appendix}

\begin{acknowledgement}
We wish to express our gratitude to the CERN accelerator division for
excellent performance of the SPS accelerator complex. We acknowledge with
appreciation the effort of all engineers, technicians and support staff 
who have participated in the construction of the experiment.

This work was supported jointly by 
the German BMBF and DFG, 
the U.S. DOE,
the Swedish NFR and FRN, 
the Dutch Stichting FOM, 
%the Stiftung f{\"u}r Deutsch-Polnische Zusammenarbeit,
Polish KBN under contract 621\-/E-78\-/SPUB\-/CERN\-/P-03\-/DZ211,
the Grant Agency of the Czech Republic under contract No. 202/95/0217,
the Department of Atomic Energy,
the Department of Science and Technology,
the Council of Scientific and Industrial Research and 
the University Grants 
Commission of the Government of India, 
the Indo-FRG Exchange Program,
the PPE division of CERN, 
the Swiss National Fund, 
% the International Science Foundation under Contract N8Y000, 
the INTAS under Contract INTAS-97-0158, 
ORISE, 
Research-in-Aid for Scientific Research
(Specially Promoted Research \& International Scientific Research)
of the Ministry of Education, Science and Culture, 
the University of Tsukuba Special Research Projects, and
the JSPS Research Fellowships for Young Scientists.
ORNL is managed by Lockheed Martin Energy Research Corporation under
contract DE-AC05-96OR22464 with the U.S. Department of Energy.
The MIT and University of Tennessee groups have been supported by the US 
Dept. of Energy under the
cooperative agreements DE-FC02-94ER40818 and DE-FG02-96ER40982, 
respectively.
\end{acknowledgement}

\end{document}